\documentclass[camera,letterpaper,nomarginnotes,nonarrowgutter]{jpaper} 

\usepackage[sort,compress]{cite}
\usepackage{amsmath,amssymb,amsfonts}
\usepackage{algorithmic}
\usepackage{graphicx}
\usepackage{textcomp}
\usepackage[bookmarks=true,breaklinks=true,letterpaper=true,colorlinks,linkcolor=black,citecolor=black,urlcolor=blue]{hyperref}
\usepackage{balance} 
\usepackage{fancyhdr}
\usepackage{datetime}
\usepackage{enumitem}
\usepackage{marginnote}
\usepackage{soul}
\usepackage{setspace}
\usepackage{dblfloatfix}    

\newcommand{\om}[1]{{#1}}
\newcommand{\atb}[1]{{#1}}
\newcommand{\ot}[1]{{#1}}
\newcommand{\omt}[1]{{#1}}
\newcommand{\of}[1]{{#1}}
\newcommand{\ofi}[1]{{#1}}
\newcommand{\agy}[1]{{#1}}
\newcommand{\secref}[1]{§\ref{#1}}

\title{A Modern Large-Scale \om{Memory} Characterization Laboratory}

\newcommand{\affilETH}[0]{\textsuperscript{\S}}

\author{
{Ataberk Olgun\affilETH}\qquad%
{Haocong Luo\affilETH}\qquad%
{\.{I}smail Emir Y\"{u}ksel\affilETH}\qquad%
{F. Nisa Bostanc\i{}\affilETH}\qquad\\
{A. Giray Ya\u{g}l\i{}kc\i{}\affilETH}\qquad
{Onur Mutlu\affilETH}\qquad\vspace{1.5mm}\\
{\affilETH SAFARI Research Group}\\
{\emph{ETH Z{\"u}rich}}}

\newcommand{\atbcomment}[1]{}
\newcommand{\otcomment}[1]{}

\newcommand{\citeAllDRAMBenderWorks}[0]{\cite{yuksel2025columndisturb,
yuksel2025pudhammer, luo2025revisiting, olgun2025variable,
tugrul2025understanding, olgun2024read, luo2023rowpress, yaglikci2024spatial,
yaglikci2022understanding, orosa2021deeper, kim2020revisiting, koppula2019eden,
khan2017detecting, kim2014flipping, liu2013experimental, baek2025marionette,
zhou2025compromising, lang2023blaster, bepary2022dram, farmani2021rhat,
yuksel2025indram, yuksel2024simultaneous, yuksel2024functionally,
yaglikci2022hira, orosa2021codic, olgun2021quac, kim2019drange, garzon2026cadm,
kubo2025pudtune, zhou2024dram, gao2022fracdram, talukder2020towards,
gao2019computedram, talukder2019prelatpuf, talukder2019exploiting,
nam2024dramscope, nam2023xray, hassan2021utrr, frigo2020trrespass,
chang2017understanding, chang2016understanding, ghose2018your,
canpolat2025easydram, luo2026dejavu, tokuda2026pudghost, tokuda2026clutch,
wang2026scaledisturb, baser2026simrapuf, yuksel2024pulsar, kim2018dram,
patel2020beer, patel2021harp, patel2019understanding, patel2017reaper,
jiang2026dram}}

\newcommand{\mitigatingRowHammerAllCitations}[0]{\cite{AppleRefInc,
rh-hp,rh-lenovo,greenfield2012throttling, kim2014flipping, kim2014architectural,
bains14d, bains14c, aweke2016anvil, bains-merged, son2017making,
seyedzadeh2018cbt,irazoqui2016mascat, you2019mrloc, lee2019twice,
park2020graphene, yaglikci2021security, yaglikci2021blockhammer,
frigo2020trrespass, kang2020cattwo, hassan2021utrr, qureshi2022hydra,
saileshwar2022randomized, brasser2017can, konoth2018zebram, van2018guardion,
vig2018rapid,  kim2022mithril, lee2021cryoguard, marazzi2022protrr,
zhang2022softtrr, joardar2022learning, juffinger2023csi, yaglikci2022hira,
saxena2022aqua, enomoto2022efficient, manzhosov2022revisiting, ajorpaz2022evax,
naseredini2022alarm, joardar2022machine, hassan2022case,
zhang2020leveraging,loughlin2021stop, devaux2021method, han2021surround,
fakhrzadehgan2022safeguard, saroiu2022price, saroiu2022configure,
loughlin2022moesiprime, zhou2022lt, hong2023dsac, mutlu2023fundamentally,
marazzi2023rega, di2023copy, sharma2022review, woo2023scalable, park2022row,
wi2023shadow, kim2023ddr5, gude2023defending, guha2022criticality,
france2022modeling, france2022reducing, bennett2021panopticon,
arikan2022processor, tomita2022extracting, saxena2023pt, zhou2023dnndefender,
woo2023rampart, kim2023how, olgun2024abacus, yaglikci2024spatial,
bostanci2024comet, saroiu2024ddr5, saxena2024start, jedecddr5c,
canpolat2024understanding,jaleel2024pride,saxena2024rubix, qureshi2026salt, 
taneja2026mirza, vittal2025mopac, kim2025per, fiedler2026memory, kim2026pvac}}

\begin{document}

\maketitle

\setstretch{1.1}

\begin{abstract}
Real \om{memory} chip characterization yields insights into fundamental
operational characteristics of modern memory, enabling new mechanisms that
improve \om{memory} performance, robustness, security, and energy efficiency. We
describe our large-scale DRAM characterization laboratory for understanding
DRAM. A key building block of this laboratory is
DRAM~Bender~\cite{olgun2023drambender, safari-drambender}, a versatile and
easy-to-use modern DRAM characterization infrastructure. We \om{have}
update\om{d} DRAM~Bender to i)~introduce support for new types of
characterization experiments, ii)~expand on its DRAM interface standard support,
and iii)~make \atb{it easier to use at large scale}. This paper introduces these
updates for the first time. \om{We hope our infrastructure enables the community
to discover new problems and solve \ot{critical memory} scaling issues, enabling the
overcoming of the huge memory bottleneck that plagues modern computing systems.}
\end{abstract}

\thispagestyle{plain}
\pagestyle{plain}

\section{Background}

\atbcomment{yanked Hasan's thesis}
DRAM~\cite{dennard1968dram} is the \ot{prevalent memory} technology
\om{overwhelmingly} used for main memory in modern systems. Unfortunately, as
DRAM scales down to smaller technology nodes, it faces key
challenges~\cite{mutlu2013memory, mutlu2025memory, mutlu2015main,
mutlu2015research, mutlu2017rowhammer, mutlu2019retrospective,
mutlu2023fundamentally} in both data integrity and latency, which strongly
affect overall system performance, robustness, security, and energy efficiency.
To develop high-performance, robust, secure, and energy-efficient DRAM-based
main memory in future systems, researchers rely on publicly-available DRAM
testing infrastructure to experimentally characterize, understand, and analyze
various aspects (e.g., reliability, latency) of \agy{real} DRAM chips. 

DRAM~Bender~\cite{olgun2023drambender, safari-drambender} is a prominent,
publicly-available DRAM testing infrastructure. \atb{This infrastructure is a
product of our \ot{almost two\agy{-}decade}\agy{-}long research into empirically
studying and understanding many interesting characteristics of modern DRAM
chips. \atbcomment{largely yanked from 2013 paper retrospective}In \ot{2011}, we
set out to rigorously understand the difficulty of DRAM data retention time
identification using an empirical approach~\cite{liu2013experimental,
mutlu2023retrospectiveretention}. No prior work at the time provided real data
on the retention characteristics of state-of-the-art DRAM chips, let alone a
detailed empirical analysis of major problems that make retention time profiling
challenging and how DRAM technology scaling affects those challenges. In fact,
\emph{no} infrastructure to study these characteristics existed (or was
available to us). We decided to build our own FPGA-based infrastructure to
characterize real DRAM chips in a flexible manner so that we could change the
refresh rate, data patterns, and other major parameters. This infrastructure,
which took us more than a year to build and which we later open sourced as
SoftMC~\cite{hassan2017softmc, softmc-safarigithub} and
DRAM~Bender~\cite{olgun2023drambender, safari-drambender}, enabled us (and
\agy{many} others) to empirically study and understand many interesting \ot{and
important} characteristics of modern DRAM chips.}

Many researchers and practitioners from industry and academia have been using
DRAM~Bender (and SoftMC) to deeply understand DRAM \om{and solve major memory
challenges}~\citeAllDRAMBenderWorks{}. \om{For example, DRAM Bender has enabled
the discovery and study of major DRAM \ot{phenomena} including
RowHammer~\cite{kim2014flipping}, RowPress~\cite{luo2023rowpress}, Variable Read
Disturbance~\cite{olgun2025variable},
ColumnDisturb~\cite{yuksel2025columndisturb}, \ot{the DejaVu
effect~\cite{luo2026dejavu}}, Variable Retention Time~\cite{qureshi2015avatar,
patel2017reaper,liu2013experimental, khan2014efficacy}, \ot{data pattern
dependence~\cite{khan2016parbor, liu2013experimental, khan2017detecting},} data
retention failures~\cite{khan2016parbor, liu2013experimental}, \ot{and
computational capabilities of real commercial off-the-shelf (COTS) DRAM
chips~\cite{yuksel2025indram, yuksel2024simultaneous, yuksel2024functionally,
yaglikci2022hira, orosa2021codic, olgun2021quac, kim2019drange, garzon2026cadm,
kubo2025pudtune, zhou2024dram, gao2022fracdram, talukder2020towards,
gao2019computedram, talukder2019prelatpuf, talukder2019exploiting,
tokuda2026pudghost, tokuda2026clutch, yuksel2024pulsar, kim2018dram,
baser2026simrapuf}}.} \atb{The earlier discoveries~\cite{kim2014flipping,
qureshi2015avatar, khan2016parbor, liu2013experimental, kim2020revisiting} are
already recognized with major awards~\cite{mutlu2023retrospectiveretention,
avatartestoftimeaward, mutlu2023retrospectiveflipping, mutlu2024jclaward,
parbortestoftimeaward, revisiting-rowhammer-toppicks} and have had major impact
on industry's thinking and handling of DRAM robustness issues.} \ot{Many read
disturbance characterization works use DRAM Bender to identify new aspects of
the vulnerability~\cite{yuksel2025columndisturb, yuksel2025pudhammer,
luo2025revisiting, olgun2025variable, tugrul2025understanding, olgun2024read,
luo2023rowpress, yaglikci2024spatial, yaglikci2022understanding,
orosa2021deeper, kim2020revisiting, baek2025marionette, zhou2025compromising,
lang2023blaster, farmani2021rhat, nam2024dramscope, luo2026dejavu,
wang2026scaledisturb}.}

\begin{figure}[!ht]
  \centering 
  \includegraphics[width=1.0\linewidth]{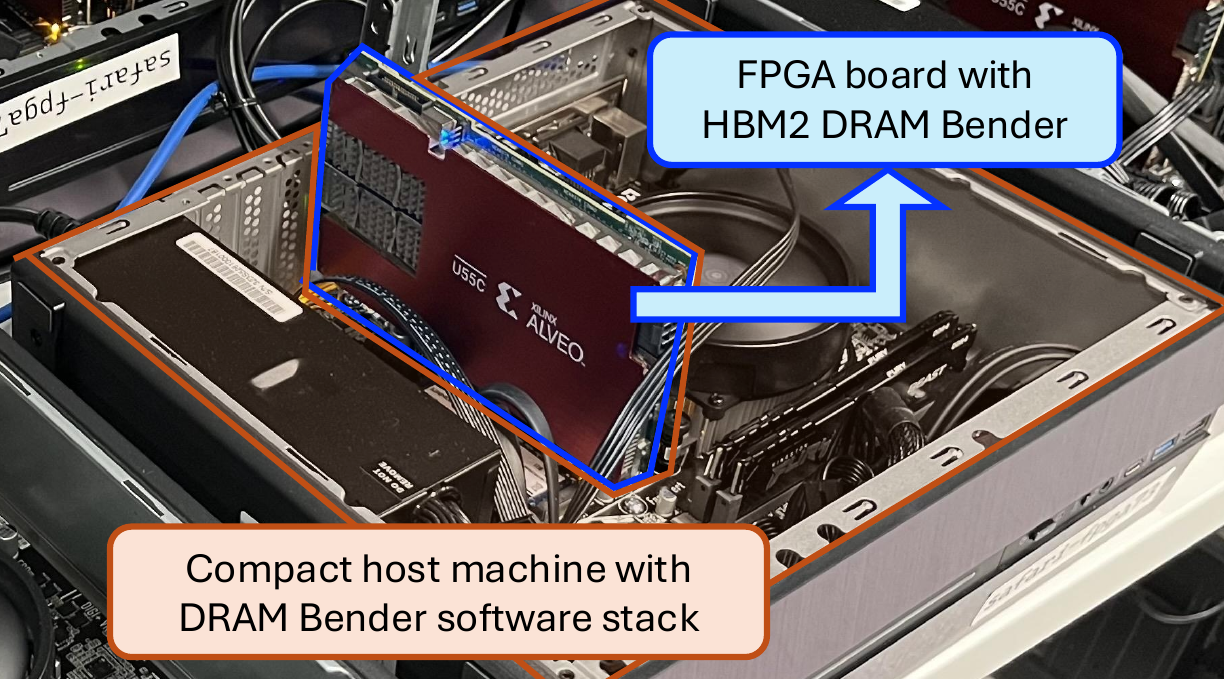}
  \caption{DRAM Bender high bandwidth memory \ot{(HBM)} characterization setup.
  \ot{More detail in~\cite{olgun2023hbm, olgun2024read, olgun2025variable}}.}
  \label{fig:dram-bender-picture}
\end{figure}

\begin{figure}[!ht]
  \centering 
  \includegraphics[width=1.0\linewidth]{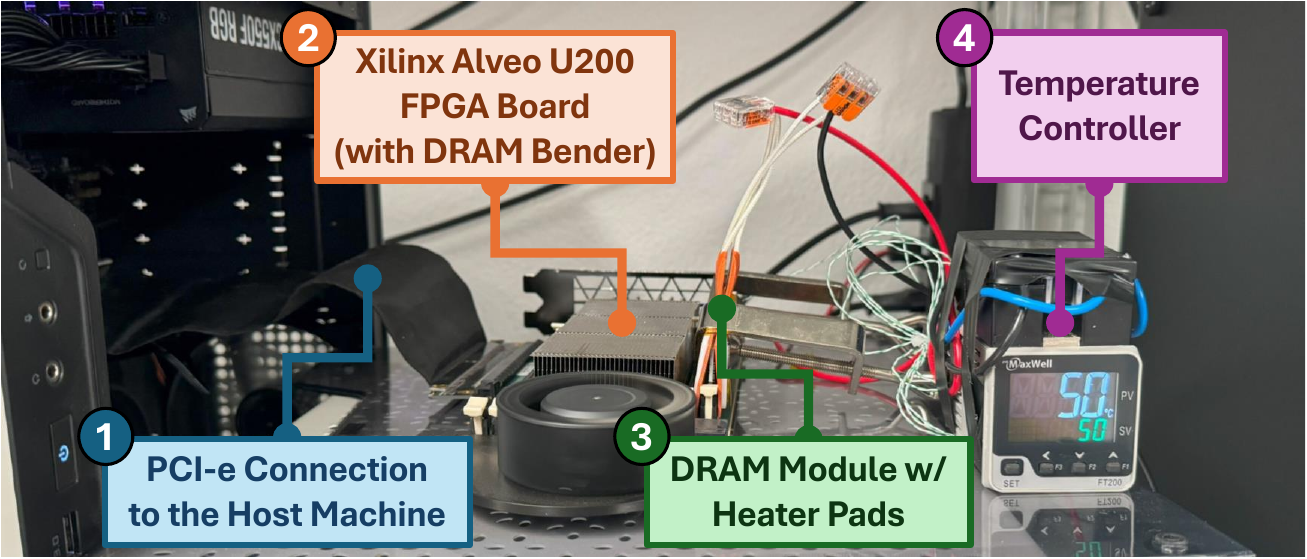}
  \caption{DRAM Bender DDR4 characterization setup}
  \label{fig:dram-bender-picture-ddr}
\end{figure}

DRAM~Bender \om{(\ot{building on its} predecessor SoftMC) has seen} widespread
use \om{due} to its three key properties: i)~flexibility in the types of DRAM
characterization experiments it supports, ii)~ease of use in a manner accessible
to both software and hardware developers, and iii)~wide availability across many
DRAM standards, DRAM module form factors, and FPGA boards. Our journal article
on DRAM~Bender~\cite{olgun2023drambender} describes how DRAM~Bender was built
from the ground up around \atb{the} three key properties.
Figure~\ref{fig:dram-bender-picture} shows a picture of a DRAM Bender setup used
to test modern 3D-stacked high bandwidth memory (HBM) stacks.
\atb{Figure~\ref{fig:dram-bender-picture-ddr} shows a picture of another DRAM
Bender setup, equipped with a temperature controller, used to test modern
\ot{DDR4} DRAM modules.} 

\section{DRAM Characterization Laboratory}

Large scale DRAM characterization accelerates the discovery of various unknown
DRAM characteristics, yielding new scientific insight to ultimately improve
DRAM-based memory system performance, robustness, security, and energy
efficiency. Access to a larger DRAM Bender characterization infrastructure
enables researchers to concurrently test more \om{new} ideas\om{, properties,}
and hypotheses, and also enables more in-depth quantitative consideration of
\om{previously-known} ideas\om{, properties,} and hypotheses that could
\omt{otherwise be} \ot{potentially} discarded due to lack of resources given a
smaller infrastructure.

\begin{figure}[!ht]
  \centering 
  \includegraphics[width=1.0\linewidth]{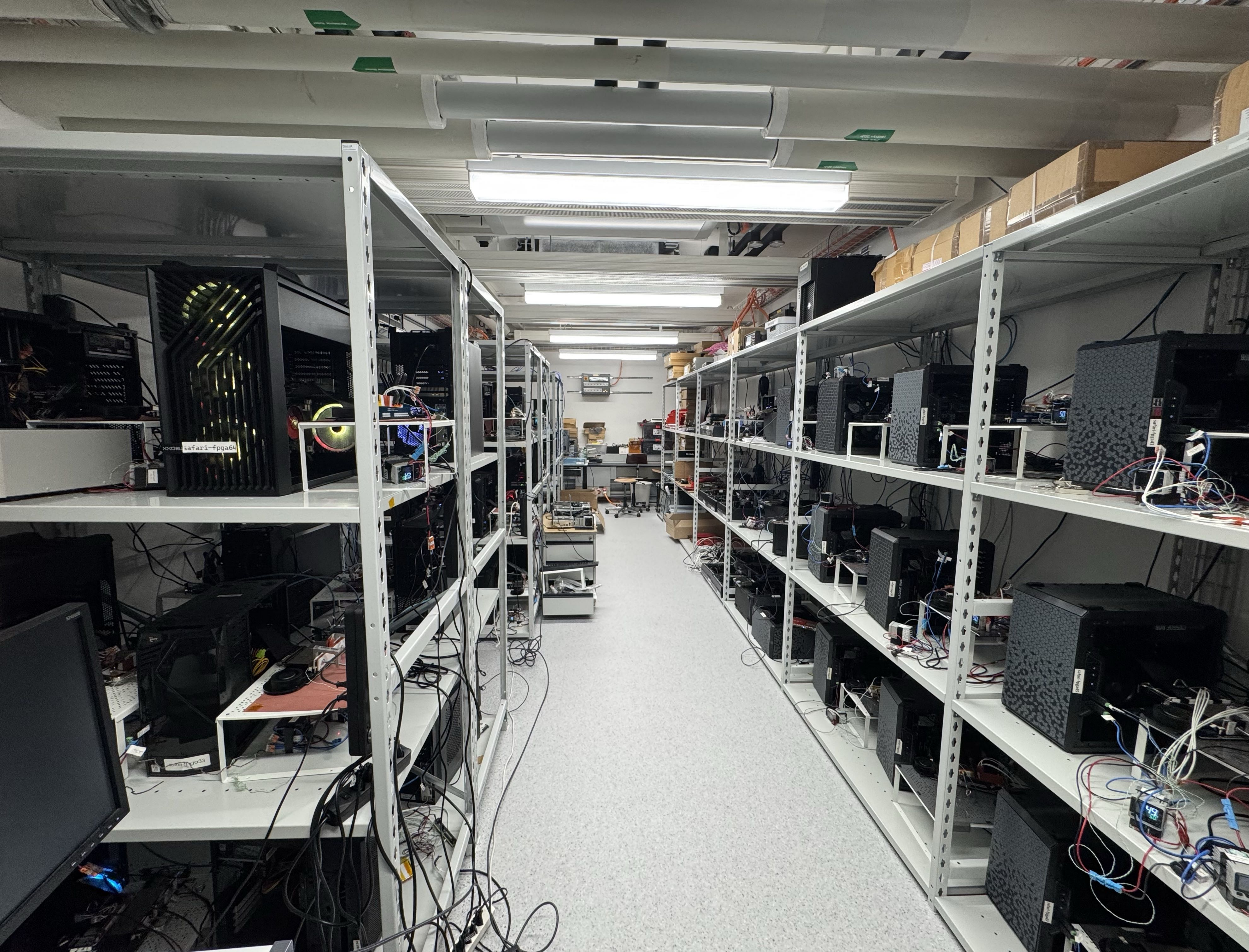}
  \caption{Our laboratory for understanding DRAM with 100+ DRAM Bender setups}
  \label{fig:dram-bender-laboratory}
\end{figure}

To accelerate scientific discovery that leads to better memory systems, we have
built a large-scale characterization laboratory comprising 100+ DRAM Bender
setups. Figure~\ref{fig:dram-bender-laboratory} shows a picture of this laboratory.

\subsection{A Versatile DRAM Infrastructure}

DRAM Bender enables a wide variety of DRAM characterization experiments. We
highlight two classes of prior works that use DRAM Bender (and its predecessor,
SoftMC) to discover new problems and solve outstanding \ot{DRAM technology}
scaling issues.

\noindent
\textbf{Understanding DRAM Robustness Problems~\cite{yuksel2025columndisturb,
yuksel2025pudhammer, luo2025revisiting, olgun2025variable,
tugrul2025understanding, olgun2024read, luo2023rowpress, yaglikci2024spatial,
yaglikci2022understanding, orosa2021deeper, kim2020revisiting, koppula2019eden,
khan2017detecting, kim2014flipping, liu2013experimental, baek2025marionette,
zhou2025compromising, lang2023blaster, bepary2022dram, farmani2021rhat,
nam2024dramscope, nam2023xray, qureshi2015avatar, tokuda2026pudghost,
tokuda2026clutch, luo2026dejavu, patel2020beer, patel2021harp,
patel2019understanding, patel2017reaper, wang2026scaledisturb, jiang2026dram,
orosa2024spyhammer}.} DRAM Bender gives its users access to the underlying
low-level DRAM interface, enabling direct observation of DRAM-device-level error
patterns. Leveraging this key feature of DRAM Bender, researchers \ot{have} made
\ot{significant} discoveries \ot{of} key DRAM robustness issues such as read
disturbance and data retention failures. The original RowHammer
paper~\cite{kim2014flipping} showed that repeatedly activating an aggressor DRAM
row many times induces bitflips in physically-neighboring victim rows. The paper
also showed that \ot{RowHammer} was a widespread phenomenon across commodity
DRAM modules \ot{manufactured} in \ot{2012-}2013. Kim \ot{et
al.}~\cite{kim2020revisiting} revisited RowHammer in 2020 by quantitatively
analyzing the RowHammer vulnerability in thousands of commodity DRAM chips and
showed that DRAM became more vulnerable to RowHammer as DRAM technology
advanced. \agy{Orosa and Yaglikci et
al.~\cite{orosa2021deeper,yaglikci2024spatial,orosa2024spyhammer} provided an
in-depth analysis of RowHammer's sensitivities to temperature, access patterns,
and physical location of victim rows.} \ot{Luo et al.}~\cite{luo2023rowpress}
discovered a new read disturbance phenomenon \ot{called RowPress,} where keeping
a DRAM row open for a very long time induces read disturbance bitflips with as
few as \emph{only} one DRAM row activation. More recently, \ot{the}
ColumnDisturb \ot{discovery}~\cite{yuksel2025columndisturb} showed that the read
disturbance \ot{effect} of activating a DRAM row is \emph{not} limited to the
few victim rows physically neighboring an aggressor row, but that the
disturbance effect can span entire \emph{subarrays} (thousands of DRAM rows)
along a DRAM column \ot{due to perturbations caused on thousands of bitlines}.
\ot{All these works} showed that \ot{these new read disturbance phenomena} can
manifest as bitflips in standard DRAM operating conditions and jeopardize the
\ot{robustness (i.e.,} security, safety, and reliability) of modern systems.
These works kickstarted an ongoing line of research and industrial development
that has led to \ot{many different types of} DRAM read disturbance mitigation
techniques~\mitigatingRowHammerAllCitations{}, \ot{including major changes to
DRAM standards in industry~\cite{jedecddr5c, canpolat2024understanding,
canpolat2025chronus}.}

\noindent
\textbf{Uncovering New DRAM Functionality~\cite{yuksel2025indram,
yuksel2024simultaneous, yuksel2024functionally, yaglikci2022hira,
orosa2021codic, olgun2021quac, kim2019drange, garzon2026cadm, kubo2025pudtune,
zhou2024dram, gao2022fracdram, talukder2020towards, gao2019computedram,
talukder2019prelatpuf, talukder2019exploiting, tokuda2026pudghost,
tokuda2026clutch, baser2026simrapuf, kim2018dram, olgun2022pidram,
yuksel2024pulsar}.} DRAM Bender enables users to issue arbitrary DRAM commands
with an arbitrary delay between subsequently issued commands to the tested DRAM
chip. Using this key feature, \ot{various works demonstrated previously-unknown
functionality in real commercial off-the-shelf (COTS) DRAM chips. \omt{The DRAM
Latency PUF~\cite{kim2018dram} showed that the DRAM read access latency can be
reduced below the reliable datasheet specifications and the resulting error
patterns used as physical unclonable functions (PUFs).}
D-RaNGe~\cite{kim2019drange} \omt{proposed a new DRAM-based true random number
generator based on the key idea of intentionally violating DRAM access timing
parameters and using the resulting errors as a source of randomness.}}
\omt{ComputeDRAM~\cite{gao2019computedram} showed that real DDR3 DRAM chips are
capable of RowClone~\cite{seshadri2013rowclone} and bitwise
majority~\cite{seshadri2017ambit, seshadri2015fast, seshadri2020indram}
operations by constructing carefully engineered DRAM command sequences with
greatly violated DRAM timing parameters and issuing these commands to DRAM chips
in quick succession.} QUAC-TRNG~\cite{olgun2021quac} showed that four DRAM rows
can be concurrently activated if timing parameters between activate and
precharge commands are greatly violated. Using this four-row activation
primitive, QUAC-TRNG generates true random numbers at higher throughput than the
state-of-the-art at the time. Building on QUAC-TRNG,
SiMRA-TRNG~\cite{yuksel2025indram} showed that even more rows can be
concurrently activated to provide even higher throughput true random number
generation. FCDRAM~\cite{yuksel2024functionally} and
SiMRA-DRAM~\cite{yuksel2024simultaneous} demonstrated that real modern COTS DRAM
chips can perform a wide range of Processing Using DRAM operations, including
functionally complete Boolean logic, many-input Boolean operations, simultaneous
many-row activation, majority operations, and multi-row copy.
\of{PiDRAM~\cite{olgun2022pidram} demonstrated COTS DRAM-based bulk data
movement and \ofi{true random number generation} in an FPGA-based prototype and showed practical benefits for bulk data
copy and initialization, while also providing foundational solutions to key COTS
DRAM-based Processing Using DRAM (PUD) integration challenges such as memory
allocation, alignment, coherence, and host-PUD coordination.}

\noindent
\textbf{Other Works~\cite{nam2024dramscope, nam2023xray, hassan2021utrr,
frigo2020trrespass, chang2017understanding, chang2016understanding,
ghose2018your, canpolat2025easydram, olgun2022pidram, gao2022fracdram,
patel2020beer, patel2021harp, patel2019understanding}.} DRAM Bender's ease of
use also enabled researchers to gain insight into DRAM
internals~\cite{nam2024dramscope, nam2023xray, hassan2021utrr,
frigo2020trrespass, patel2020beer, patel2021harp, patel2019understanding},
\ot{uncover other DRAM functionality~\cite{gao2022fracdram},} understand DRAM
voltage and latency sensitivity~\cite{chang2017understanding,
chang2016understanding}, characterize DRAM power
\ot{consumption}~\cite{ghose2018your}. Part of DRAM Bender's hardware design was
reused in other system evaluation infrastructure~\cite{canpolat2025easydram,
olgun2022pidram}.

\section{New DRAM Bender Features}

We have updated DRAM Bender's hardware and software components with new features
to 1)~\ot{enable} DRAM Bender \ot{to} support new types of experiments
\ot{(\secref{sec:dram-power-measurement})}, 2)~expand on its DRAM interface
standard support \ot{(\secref{sec:hbm2})}, and 3)~make it \atb{easier to use at
large scale} \ot{(\secref{sec:slurm})}. We \ot{strongly} believe these features
will accelerate discovery \ot{and experimental studies} that \ot{would lead} to
better memory \ot{and DRAM-based computing} systems.

\subsection{DRAM Power Measurement}
\label{sec:dram-power-measurement}

\atbcomment{yanked from our under submission hbm2 power paper}Power consumption
of main memory is a critical system design constraint. Unfortunately, detailed
empirical power consumption data for newer DRAM standards is scarce. Our updated
infrastructure provides DRAM power measurement support for DDR4 modules and for
HBM2 stacks, enabling users to empirically measure DRAM power using commodity
DRAM and develop better power models. We aim to publicly release \ot{our new
validated power models and} the sources for our power measurement setups after
we rigorously test and calibrate them \ot{(as we have done in the past for DDR3
DRAM~\cite{ghose2018your, safari-vampire})}.

\begin{figure}[!ht]
  \centering 
  \includegraphics[width=1.0\linewidth]{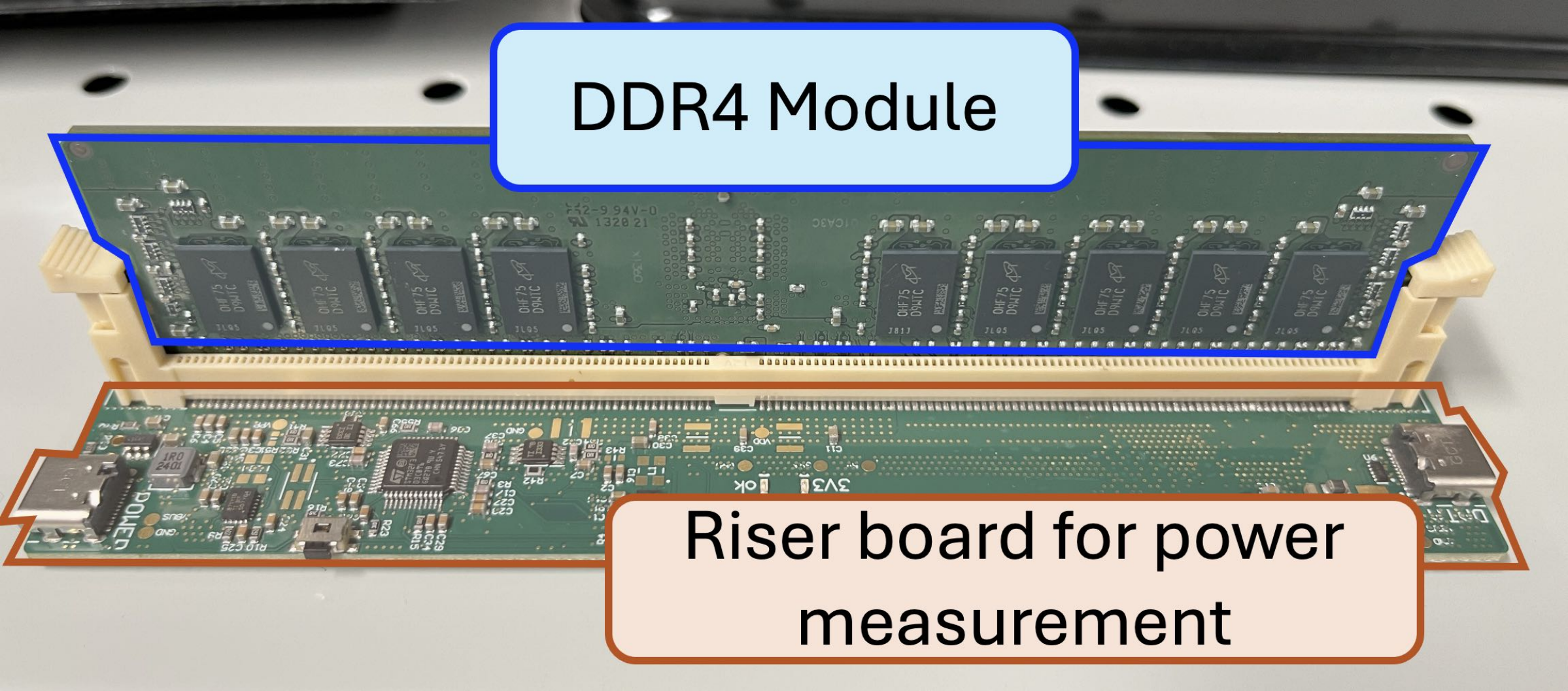}
  \caption{Custom DDR4 \ot{DRAM} riser board used for power measurement}
  \label{fig:dram-bender-power-riser}
\end{figure}

\noindent
\textbf{DDR4 Power Measurement.} To enable fine-grained power measurement of
DDR4 modules in DRAM Bender, we designed a dedicated riser board that sits
between the DIMM slot and the DDR4 module.
Figure~\ref{fig:dram-bender-power-riser} shows a picture of the riser board. The
board intercepts the power rails delivered to the module and routes these rails
through on-board components to provide accurate, real-time power readings. The
board connects to a host machine via a universal serial bus (USB) link. We use
the USB link to record power data during DRAM Bender experiments. 

\noindent
\textbf{HBM2 Power Measurement.} We provide fine-grained power measurement
support for HBM2 stacks in DRAM Bender by integrating the AMD Card Management
Subsystem (CMS)~\cite{amd-alveo-cms-manual} into an Alveo
U55C~\cite{alveo-u55-c} (HBM2-based FPGA board) DRAM Bender
setup.\footnote{\agy{Any FPGA board that supports
CMS~\cite{amd-alveo-cms-manual} and provides current measurement capability on
DRAM power rails (e.g., Alveo U55C~\cite{alveo-u55-c}) can be used with DRAM
Bender to conduct power measurement studies.}} CMS periodically queries an on
board satellite controller for current measured on the primary HBM2 voltage
rail. We augment the DRAM Bender software stack with a routine that fetches
measured instantenous, maximum, and average current values via the PCIe
interface from the DRAM Bender setup.

\subsection{High Bandwidth Memory (HBM2) Support}
\label{sec:hbm2}

\begin{figure}[!ht]
  \centering 
  \includegraphics[width=1.0\linewidth]{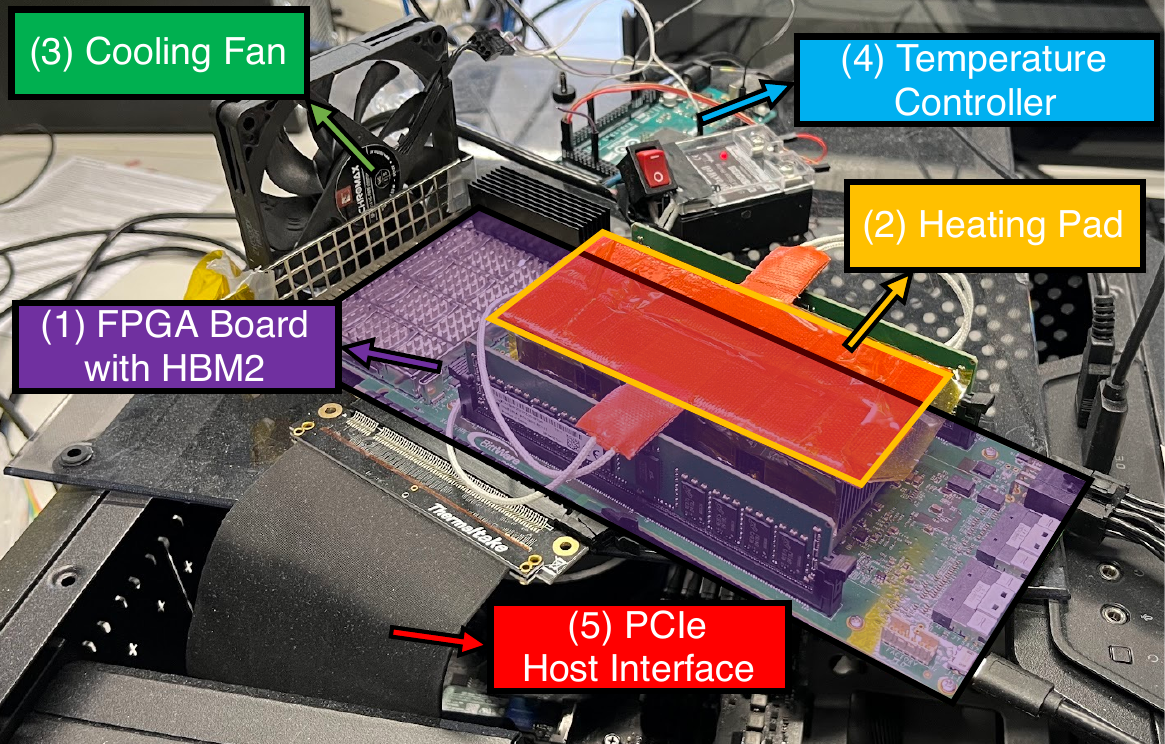}
  \caption{HBM2 DRAM Bender setup as depicted in~\cite{olgun2024read}}
  \label{fig:hbm2-dram-bender}
\end{figure}

To meet the high-bandwidth requirements of modern data-intensive
applications~\cite{bakhoda2009analyzing, che2009rodinia, jouppi2023tpuv4,
brown2020language, devlin2019bert, boroumand2018google, boroumand2021google,
ahn2015scalable, oliveira2021damov, lee2016simultaneous}, DRAM designers develop
HBM stacks, which contain multiple layers of 3D-stacked DRAM dies. We have
updated our infrastructure with a new DRAM Bender design that supports HBM2 in
the Bittware XUPVVH~\cite{xupvvh} and Alveo U50~\cite{alveo-u50} FPGA
boards.\footnote{\agy{We have ported DRAM Bender to AMD and Bittware boards so
far, but there is no fundamental limitation that prevents DRAM Bender \ofi{from being}
ported to other FPGA boards from other manufacturers or custom FPGA boards.}}
These designs are already publicly available in our GitHub
repository~\cite{safari-drambender}. Figure~\ref{fig:hbm2-dram-bender} shows a
picture of one of our HBM2-based DRAM Bender setups.

DRAM Bender allows precise control of the HBM2 command timings at the
granularity of 1.67 ns (i.e., the HBM2 interface clock speed is 600 MHz). Each
\atbcomment{Double check} Bittware XUPVVH and Alveo U50 board has a 4 GiB HBM2
stack with 8 channels, 2 pseudo channels, 16 banks, 16K rows, and 1 KiB row
size. We used the HBM2-based DRAM Bender design to show for the first time that
read disturbance (RowHammer and RowPress) was prevalent across real HBM2
stacks~\cite{olgun2024read, olgun2025variable, olgun2023hbm}. 

\subsection{Job Scheduling System for Large-Scale Experiments}
\label{sec:slurm}

Experimental rigor in hypothesis testing requires gathering characterization
data from a wide variety of DRAM chips spanning different chip organizations,
manufacturers, die revisions, and die densities. Without access to a large-scale
laboratory, researchers must physically replace DRAM modules in a small number
of DRAM Bender setups as they conduct experiments across a diverse pool of DRAM
chips. This manual process i)~is labor-intensive, ii)~\ot{gradually damages}
DRAM module slots on FPGA boards \ot{to a point of failure}, and
iii)~accelerates the aging of DRAM Bender setup components because of repeated
\ot{manual} handling. Our large-scale laboratory eliminates this overhead by
\emph{statically allocating} a wide variety of DRAM chips across our 100+ DRAM
Bender setups: each DRAM module is installed once and rarely replaced with
another module setup, such that no manual DRAM module replacement is required
during typical use. 

Static allocation introduces new resource allocation problems. First, each
submitted experiment must be dispatched to a setup that hosts a DRAM chip
matching the experiment's requirements (e.g., a target manufacturer, die
revision, or DRAM standard). Second, contention between concurrent experiments
must be resolved fairly. We address this problem by integrating DRAM Bender with
Slurm~\cite{yoo2003slurm}, a widely-used open-source workload
manager\otcomment{correct?}. We model each DRAM Bender setup as a Slurm node and
expose the attributes of the setup's installed DRAM chip and supplementary
hardware (e.g., temperature controller, power measurement board) as Slurm
features. Researchers submit experiments as Slurm jobs tagged with the required
features, and Slurm dispatches each job to a compatible idle setup as resources
become available. We aim to publicly release our Slurm integration alongside
DRAM Bender to enable other research groups to easily operate similar
large-scale characterization laboratories.

\section{DRAM Bender Adoption \& Community}

DRAM Bender has been adopted by research groups across both industry and
academia~\citeAllDRAMBenderWorks{}, yielding a growing body of work to discover
new problems \omt{and} \ot{new capabilit\omt{ies},} and solve outstanding
\ot{DRAM technology} scaling issues. To further accelerate adoption and lower
the barrier to entry for new researchers, we organize a recurring joint
Ramulator~\cite{ramulator2github, ramulatorgithub, kim2016ramulator,
luo2024ramulator, bostanci2026cleaning} and DRAM Bender tutorial series co-located with major computer
architecture conferences. Each tutorial provides an extensive overview of both
i) Ramulator, our cycle-accurate and extensible main memory simulator, and ii)
DRAM Bender, covering its programming interface, supported DRAM standards, and
example characterization workflows. The tutorial series also serves as a venue
for the community to present and discuss new research conducted using DRAM
Bender. We have so far successfully held the first edition of the tutorial at
ASPLOS 2026~\cite{tutorial-asplos26}, which featured invited presentations on
DRAM characterization and memory system simulation. We are organizing the second
edition at ISCA 2026~\cite{tutorial-isca26}, where we plan to additionally
provide a hands-on session that walks attendees through writing, running, and
analyzing the results of real DRAM characterization experiments using DRAM
Bender. We are also organizing a third edition at ICS
2026~\cite{tutorial-ics26}, whose workshop proceedings include this paper. 

\agy{We offer the DRAM Bender projects \& seminars
course~\cite{psdb-latest-edition, psdb-2024} where ETH Z\"{u}rich undergraduate
students learn how DRAM is organized and operates in a low-level and gain
practical experience in using DRAM Bender. The sources for this course are
available online (see~\cite{psdb-latest-edition} for the latest edition) and the
recordings of the lectures are uploaded to YouTube for Spring
2022~\cite{psdb-spring2022}, Fall 2022~\cite{psdb-fall2022}, Spring
2023~\cite{psdb-spring2023}, Fall 2023~\cite{psdb-2024}, and Spring
2024~\cite{psdb-2024} editions. These playlists feature DRAM Bender tutorials
and research talks on scientific papers that use DRAM Bender.}

We hope our tutorial series\agy{, course offerings}, and the continued
open-source development of DRAM Bender will enable a broader segment of the
community to conduct rigorous experimental DRAM research \ot{and discover more
new characteristics and ideas in memory systems}.

\section*{Acknowledgments}
We thank the SAFARI Research Group members for their constructive feedback
\ot{for decades} and for providing a stimulating intellectual, scientific\ot{,
scholarly} environment. We acknowledge the generous gift funding provided by our
industrial partners (especially Google, Huawei, Intel, Microsoft), which has
been instrumental in enabling the research we have been conducting on read
disturbance in DRAM in particular and memory systems in
general~\cite{mutlu2013memory, mutlu2025memory, mutlu2024memory,
mutlu2023fundamentally, mutlu2025modern, mutlu2017rowhammer,
mutlu2019retrospective, mutlu2019processing, mutlu2015research,
mutlu2023retrospectiveretention, ghose2019processing,
mutlu2023retrospectiveflipping, mutlu2015main, luna2022benchmarking,
oliveira2022accelerating, singh2021fpga, oliveira2021damov,
kakolyris2026columnkeeper, cai2017flashtbd, mutlu2020intelligentdate,
mutlu2023tesseractretrospective, mutlu2023selfretrospective,
luo2026ramulator2.1, yuksel2026memory, bostanci2026cutmess, cai2017errors}. This
work was in part supported by a Google Security and Privacy Research Award and
the Microsoft Swiss Joint Research Center.

\setstretch{1.0}
\balance
\bibliographystyle{unsrt}
\bibliography{refs}

@inproceedings{olgun2025variable,
	title={{Variable Read Disturbance: An Experimental Analysis of Temporal Variation in DRAM Read Disturbance}},
	author={Olgun, Ataberk and Bostanc{\i}, F Nisa and Y{\"u}ksel, {\.I}smail Emir and Canpolat, O{\u{g}}uzhan and Luo, Haocong and Oliveira, Geraldo F and Ya{\u{g}}l{\i}k{\c{c}}{\i}, A Giray and Patel, Minesh and Mutlu, Onur},
	booktitle={HPCA},
	year={2025}
}

@inproceedings{yoo2003slurm,
  title={{Slurm: Simple Linux Utility for Resource Management}},
  author={Yoo, Andy B and Jette, Morris A and Grondona, Mark},
  booktitle={JSSPP},
  year={2003}
}

@inproceedings{patel2019understanding,
  author={Patel, Minesh and Kim, Jeremie S. and Hassan, Hasan and Mutlu, Onur},
  booktitle={DSN}, 
  title={{Understanding and Modeling On-Die Error Correction in Modern DRAM: An Experimental Study Using Real Devices}}, 
  year={2019}
}

@inproceedings{kim2026pvac,
      title={{PVAC: A RowHammer Mitigation Architecture Exploiting Per-victim-row Counting}}, 
      author={Jumin Kim and Seungmin Baek and Hwayong Nam and Minbok Wi and Nam Sung Kim and Jung Ho Ahn},
      year={2026},
			booktitle={ISCA}
}

@inproceedings{seshadri2017ambit,
  title     = {{Ambit: In-Memory Accelerator for Bulk Bitwise Operations Using Commodity DRAM Technology}},
  author    = {Seshadri, Vivek and Lee, Donghyuk and Mullins, Thomas and Hassan, Hasan and Boroumand, Amirali and Kim, Jeremie and Kozuch, Michael A. and Mutlu, Onur and Gibbons, Phillip B. and Mowry, Todd C.},
  booktitle = {MICRO},
  year      = {2017}
}

@inproceedings{boroumand2018google,
  title={{Google Workloads for Consumer Devices: Mitigating Data Movement Bottlenecks}},
  author={Boroumand, Amirali and Ghose, Saugata and Kim, Youngsok and Ausavarungnirun, Rachata and Shiu, Eric and Thakur, Rahul and Kim, Daehyun and Kuusela, Aki and Knies, Allan and Ranganathan, Parthasarathy and others},
  booktitle={ASPLOS},
  year={2018},
}

@inproceedings{boroumand2021google,
  title={{Google Neural Network Models for Edge Devices: Analyzing and Mitigating Machine Learning Inference Bottlenecks}},
  author={Boroumand, Amirali and Ghose, Saugata and Akin, Berkin and Narayanaswami, Ravi and Oliveira, Geraldo F and Ma, Xiaoyu and Shiu, Eric and Mutlu, Onur},
  booktitle={PACT},
  year={2021},
}

@inproceedings{ahn2015scalable,
 author = {Ahn, Junwhan and Hong, Sungpack and Yoo, Sungjoo and Mutlu, Onur and Choi, Kiyoung},
 title = {{A Scalable Processing-in-Memory Accelerator for Parallel Graph Processing}},
 booktitle = {ISCA},
 year = {2015},
}

@inproceedings{jiang2026dram,
author = {Jiang, Yuelin and Lu, Bin and Wang, Xinhe and Li, Xujing and Yan, Yinan and Liu, Mingli and Bai, Zhaoqiang and Zhang, Jing and Pan, Jinlong and Huo, Qiang and Li, Yi and Dai, Jin and Kang, Bryan and Tang, Jianshi and Wu, Huaqiang and Wang, Guilei and Zhao, Chao and Cao, Kanyu},
title = {{DRAM Operations under Cryogenic Temperatures: From Device Physics to DIMM Behavior}},
year = {2026},
booktitle = {IMW}
}

@misc{seshadri2020indram,
	title={{In-DRAM Bulk Bitwise Execution Engine}}, 
	author={Vivek Seshadri and Onur Mutlu},
	booktitle={Invited Book Chapter in {Advances in Computers}},   
	year={2020}
}

@article{seshadri2015fast,
  title={{Fast Bulk Bitwise AND and OR in DRAM}},
  author={Seshadri, Vivek and Hsieh, Kevin and Boroum, Amirali and Lee, Donghyuk and Kozuch, Michael A and Mutlu, Onur and Gibbons, Phillip B and Mowry, Todd C},
  journal={IEEE Computer Architecture Letters},
  year={2015},
}

@article{oliveira2021damov,
  title={{DAMOV: A New Methodology and Benchmark Suite for Evaluating Data Movement Bottlenecks}}, 
  author={Oliveira, Geraldo F. and Gómez-Luna, Juan and Orosa, Lois and Ghose, Saugata and Vijaykumar, Nandita and Fernandez, Ivan and Sadrosadati, Mohammad and Mutlu, Onur},
  journal = {IEEE Access},
  year={2021}
}

@article{lee2016simultaneous,
  title={{Simultaneous Multi-Layer Access: Improving 3D-Stacked Memory Bandwidth at Low Cost}},
  author={Lee, Donghyuk and Ghose, Saugata and Pekhimenko, Gennady and Khan, Samira and Mutlu, Onur},
  journal={TACO},
  year={2016},
}

@article{kim2016ramulator,
author = {Kim, Yoongu and others},
journal = {CAL},
title = {{Ramulator: A Fast and Extensible DRAM Simulator}},
year = {2016}
}

@misc{ramulatorgithub,
  author = "{SAFARI Research Group}",
  title = {{Ramulator --- GitHub Repository}},
  howpublished  = {\url{https://github.com/CMU-SAFARI/ramulator}},
  year={2021}
}

@misc{tutorial-asplos26,
  author = "{SAFARI Research Group}",
  title = {{Ramulator \& DRAM Bender Tutorial at ASPLOS 2026}},
  howpublished  = {\url{https://events.safari.ethz.ch/asplos26-ramulator-drambender/}},
  year={2026}
}

@misc{tutorial-isca26,
  author = "{SAFARI Research Group}",
  title = {{2nd Ramulator \& DRAM Bender Tutorial at ISCA 2026}},
  howpublished  = {\url{https://events.safari.ethz.ch/isca26-ramulator-drambender/}},
  year={2026}
}

@misc{tutorial-ics26,
  author = "{SAFARI Research Group}",
  title = {{3rd Ramulator \& DRAM Bender Tutorial at ICS 2026}},
  howpublished  = {\url{https://events.safari.ethz.ch/ics26-ramulator-drambender/}},
  year={2026}
}

@misc{ramulator2github,
  author = "{SAFARI Research Group}",
  title = {{Ramulator 2.0 --- GitHub Repository}},
  howpublished  = {\url{https://github.com/CMU-SAFARI/ramulator2}},
  year={2023}
}

@ARTICLE{luo2024ramulator,
author = {Haocong Luo and Yahya Can Tugrul and F. Nisa Bostanci and Ataberk Olgun and Abdullah Giray Yaglikci and Onur Mutlu},
journal = {IEEE CAL},
title = {{Ramulator 2.0: A Modern, Modular, and Extensible DRAM Simulator}},
year = {2024}
}

@inproceedings{olgun2024read,
	title        = {{Read Disturbance in High Bandwidth Memory: A Detailed Experimental Study on HBM2 DRAM Chips}},
	author       = {Ataberk Olgun and Majd Osseiran and Abdullah Giray Yaglikci and Yahya Can Tugrul and Haocong Luo and Steve Rhyner and Behzad Salami and Juan Gomez Luna and Onur Mutlu},
	year         = 2024,
	booktitle    = {DSN}
}

@inproceedings{olgun2024abacus,
    title={{ABACuS: All-Bank Activation Counters for Scalable and Low Overhead RowHammer Mitigation}}, 
    author={Ataberk Olgun and Yahya Can Tugrul and Nisa Bostanci and Ismail Emir Yuksel and Haocong Luo and Steve Rhyner and Abdullah Giray Yaglikci and Geraldo F. Oliveira and Onur Mutlu},
    year={2024},
    booktitle={USENIX Security}
     
}

@inproceedings{lang2023blaster,
  title={{BLASTER: Characterizing the Blast Radius of Rowhammer}},
  author={Lang, Zhenrong and Jattke, Patrick and Marazzi, Michele and Razavi, Kaveh},
  booktitle={DRAMSec},
  year={2023}
}

@inproceedings{kim2014flipping,
	title        = {{Flipping Bits in Memory Without Accessing Them: An Experimental Study of DRAM Disturbance Errors}},
	author       = {Y. {Kim} and R. {Daly} and J. {Kim} and C. {Fallin} and J. H. {Lee} and D. {Lee} and C. {Wilkerson} and K. {Lai} and O. {Mutlu}},
	year         = 2014,
	booktitle    = {ISCA}
}

@inproceedings{bostanci2024comet,
    title={{CoMeT: Count-Min-Sketch-based Row Tracking to Mitigate RowHammer at Low Cost}},
    author={Bostanci, F Nisa and Y{\"u}ksel, ISmail Emir and Olgun, Ataberk and Kanellopoulos, Konstantinos and Tu{\u{g}}rul, Yahya Can and Ya{\u{g}}li{\c{c}}i, A Giray and Sadrosadati, Mohammad and Mutlu, Onur},
    booktitle={HPCA},
    year={2024}
}

@inproceedings{yaglikci2024spatial,
  title = {{Spatial Variation-Aware Read Disturbance Defenses: Experimental Analysis of Real DRAM Chips and Implications on Future Solutions}},
  author = {Ya{\u{g}}l{\i}k{c}{\i}, A. Giray and Tu{\u{g}}rul, Yahya Can and De Oliviera, Geraldo F and Yüksel, Ismail Emir and Olgun, Ataberk and Luo, Haocong and Mutlu, Onur},
  booktitle = {HPCA},
  year = {2024},
}

@misc{alveo-u50,
    author={{AMD Xilinx}},
    title = {{AMD Xilinx Alveo U50 FPGA Board}},
    howpublished={\url{https://www.amd.com/en/products/accelerators/alveo/u50/a-u50-p00g-pq-g.html}}
}

@article{cai2017flashtbd,
	title        = {{Error Characterization, Mitigation, and Recovery in Flash Memory Based Solid-State Drives}},
	author       = {Y. Cai and S. Ghose and E. F. Haratsch and Y. Luo and O. Mutlu},
	year         = 2017,
	journal      = {Proc. IEEE}
}

@inproceedings{mutlu2020intelligentdate,
	title        = {{Intelligent Architectures for Intelligent Computing Systems}},
	author       = {Mutlu, Onur},
	year         = 2021,
	booktitle    = {DATE}
}

@misc{mutlu2023tesseractretrospective,
  title={{Retrospective: A Scalable Processing-in-Memory Accelerator for Parallel Graph Processing}},
  author={Ahn, Junwhan and Hong, Sungpack and Yoo, Sungjoo and Mutlu, Onur and Choi, Kiyoung},
  howpublished={Retrospective Issue for ISCA-50},
  year={2023}
}

@misc{mutlu2023selfretrospective,
  title={{Retrospective: Self-optimizing Memory Controllers: A Reinforcement Learning Approach}},
  author={Jos{\'e} F. Mart{\'\i}nez and Engin Ipek and Onur Mutlu and Rich Caruana},
  howpublished={Retrospective Issue for ISCA-50},
  year={2023}
}

@misc{luo2026ramulator2.1,
  title={{Ramulator 2.1: A Composable Memory System Simulator for Modern DRAM Systems}},
  author={Haocong Luo and F. Nisa Bostancı and Ataberk Olgun and Maria Makeenkova and Ziad Malik and Ipek Akdeniz and Onur Mutlu},
  howpublished={The 3rd Tutorial on Ramulator and DRAM Bender colocated with ICS},
  year={2026}
}

@misc{yuksel2026memory,
  title={{Memory-Centric Computing: Security Benefits and Challenges of Processing-in-DRAM}},
  author={Ismail Emir Yuksel and F. Nisa Bostanci and Ataberk Olgun and Onur Mutlu},
  howpublished={The 7th Workshop on Memory-Centric Computing Systems colocated with ICS},
  year={2026}
}

@misc{bostanci2026cutmess,
  title={{Extended Abstract: Re-Evaluating the Real-System Modeling Accuracy of Ramulator 2.0}},
  author={F. Nisa Bostanci and Haocong Luo and Ataberk Olgun and Maria Makeenkova and Geraldo F. De Oliveira and A. Giray Yaglikci and Onur Mutlu},
  howpublished={The 3rd Tutorial on Ramulator and DRAM Bender colocated with ICS},
  year={2026}
}

@article{cai2017errors,
  title={{Errors in Flash-Memory-Based Solid-State Drives: Analysis, Mitigation, and Recovery}},
  author={Cai, Yu and Ghose, Saugata and Haratsch, Erich F and Luo, Yixin and Mutlu, Onur},
  journal={arXiv preprint arXiv:1711.11427},
  year={2017}
}

@misc{saroiu2024ddr5, 
    author={Saroiu, Stefan},
    title={{DDR5 Spec Update Has All It Needs to End Rowhammer: Will It?}},
    howpublished={\url{https://stefan.t8k2.com/rh/PRAC/index.html}}
}

@inproceedings{saxena2024start,
    author={Saxena, Anish and Qureshi, Moinuddin},
    booktitle={HPCA}, 
    title={{START: Scalable Tracking for any Rowhammer Threshold}}, 
    year={2024}
}

@inproceedings{brasser2017can,
	title        = {{Can't Touch This: Software-Only Mitigation Against Rowhammer Attacks Targeting Kernel Memory}},
	author       = {Brasser, Ferdinand and Davi, Lucas and Gens, David and Liebchen, Christopher and Sadeghi, Ahmad-Reza},
	year         = 2017,
	booktitle    = {USENIX Security}
}

@inproceedings{mutlu2017rowhammer,
	title        = {{The RowHammer Problem and Other Issues We May Face as Memory Becomes Denser}},
	author       = {Mutlu, Onur},
	year         = 2017,
	booktitle    = {DATE}
}

@inproceedings{van2018guardion,
	title        = {{GuardION: Practical Mitigation of DMA-Based Rowhammer Attacks on ARM}},
	author       = {van der Veen, Victor and Lindorfer, Martina and Fratantonio, Yanick and Pillai, Harikrishnan Padmanabha and Vigna, Giovanni and Kruegel, Christopher and Bos, Herbert and Razavi, Kaveh},
	year         = 2018,
	booktitle    = {DIMVA}
}

@inproceedings{frigo2020trrespass,
	title        = {{TRRespass: Exploiting the Many Sides of Target Row Refresh}},
	author       = {Frigo, Pietro and Vannacci, Emanuele and Hassan, Hasan and {van der Veen}, Victor and Mutlu, Onur and Giuffrida, Cristiano and Bos, Herbert and Razavi, Kaveh},
	year         = 2020,
	booktitle    = {{S\&P}}
}

@inproceedings{hassan2021utrr,
	title        = {{Uncovering In-DRAM RowHammer Protection Mechanisms: A New Methodology, Custom RowHammer Patterns, and Implications}},
	author       = {Hassan, Hasan and Tugrul, Yahya Can and Kim, Jeremie S. and van der Veen, Victor and Razavi, Kaveh and Mutlu, Onur},
	year         = 2021,
	booktitle    = {MICRO}
}

@article{orosa2024spyhammer,
	title        = {{SpyHammer: Understanding and Exploiting RowHammer Under Fine-Grained Temperature Variations}},
	author       = {Orosa, Lois and R{\"u}hrmair, Ulrich and Yaglikci, A Giray and Luo, Haocong and Olgun, Ataberk and Jattke, Patrick and Patel, Minesh and Kim, Jeremie and Razavi, Kaveh and Mutlu, Onur},
	year         = 2024,
	journal = {IEEE Access}
}

@inproceedings{mutlu2024memory,
  title={{Memory-Centric Computing: Recent Advances in Processing-in-DRAM}},
  author={Mutlu, Onur and Olgun, Ataberk and Oliveira, Geraldo F and Yuksel, Ismail E},
  booktitle={IEDM},
  year={2024}
}

@inproceedings{kim2020revisiting,
	title        = {{Revisiting RowHammer: An Experimental Analysis of Modern Devices and Mitigation Techniques}},
	author       = {Kim, Jeremie S. and Patel, Minesh and Ya\u{g}l{\i}kc{\i}, Abdullah Giray and Hassan, Hasan and Azizi, Roknoddin and Orosa, Lois and Mutlu, Onur},
	year         = 2020,
	booktitle    = {ISCA}
}

@inproceedings{orosa2021deeper,
	title        = {{A Deeper Look into RowHammer's Sensitivities: Experimental Analysis of Real DRAM Chips and Implications on Future Attacks and Defenses}},
	author       = {Orosa, Lois and Ya{\u{g}}l{\i}k{{c}}{\i}, A Giray and Luo, Haocong and Olgun, Ataberk and Park, Jisung and Hassan, Hasan and Patel, Minesh and Kim, Jeremie S. and Mutlu, Onur},
	year         = 2021,
	booktitle    = {MICRO}
}

@inproceedings{yaglikci2022understanding,
	title        = {{Understanding RowHammer Under Reduced Wordline Voltage: An Experimental Study Using Real DRAM Devices}},
	author       = {Ya{\u{g}}l{\i}k{c}{\i}, A. Giray and Luo, Haocong and De Oliviera, Geraldo F and Olgun, Ataberk and Patel, Minesh and Park, Jisung and Hassan, Hasan and Kim, Jeremie S and Orosa, Lois and Mutlu, Onur},
	year         = 2022,
	booktitle    = {DSN}
}

@inproceedings{mutlu2023fundamentally,
	title        = {{Fundamentally Understanding and Solving RowHammer}},
	author       = {Mutlu, Onur and Olgun, Ataberk and Yaglikci, A. Giray},
	year         = 2023,
	booktitle    = {ASP-DAC}
}

@inproceedings{olgun2021quac,
    title={{QUAC-TRNG: High-Throughput True Random Number Generation Using Quadruple Row Activation in Commodity DRAM Chips}},
    author={Olgun, Ataberk and Patel, Minesh and Ya{\u{g}}l{\i}k{\c{c}}{\i}, A Giray and Luo, Haocong and Kim, Jeremie S and Bostanc{\i}, F Nisa and Vijaykumar, Nandita and Ergin, O{\u{g}}uz and Mutlu, Onur},
    booktitle={ISCA},
    year={2021}
}

@inproceedings{kim2019drange,
    title={{D-RaNGe: Using Commodity DRAM Devices to Generate True Random Numbers with Low Latency and High Throughput}},
    author={Kim, Jeremie S and Patel, Minesh and Hassan, Hasan and Orosa, Lois and Mutlu, Onur},
    booktitle={HPCA},
    year={2019}
}

@article{garzon2026cadm,
    title={{CADM: Content Addressable Commodity Off-the-Shelf DRAM-based Genome Classifier}},
    author={Garz{\'o}n, Esteban and Fish, Alexander and Yavits, Leonid},
    journal={Journal of Systems Architecture},
    year={2026}
}

@inproceedings{zhou2024dram,
    title={{DRAM-Locker: A General-Purpose DRAM Protection Mechanism Against Adversarial DNN Weight Attacks}},
    author={Zhou, Ranyang and Ahmed, Sabbir and Roohi, Arman and Rakin, Adnan Siraj and Angizi, Shaahin},
    booktitle={DATE},
    year={2024}
}

@article{kubo2025pudtune,
    title={{PUDTune: Multi-Level Charging for High-Precision Calibration in Processing-Using-DRAM}},
    author={Kubo, Tatsuya and Tokuda, Daichi and Qu, Lei and Cao, Ting and Takamaeda-Yamazaki, Shinya},
    journal={IEEE CAL},
    year={2025}
}

@inproceedings{gao2022fracdram,
  title={{FracDRAM: Fractional Values in Off-the-Shelf DRAM}},
  author={Gao, Fei and Tziantzioulis, Georgios and Wentzlaff, David},
  booktitle={MICRO},
  year={2022}
}

@inproceedings{olgun2023hbm,
	title        = {{An Experimental Analysis of RowHammer in HBM2 DRAM Chips}},
	author       = {Olgun, Ataberk and Osseiran, Majd and Yaglikci, Abdullah Giray and Tugrul, Yahya Can and Luo, Haocong and Rhyner, Steve and Salami, Behzad and Gomez Luna, Juan and Mutlu, Onur},
	year         = 2023,
	booktitle    = {DSN Disrupt}
}

@article{olgun2023drambender,
	title        = {{DRAM Bender: An Extensible and Versatile FPGA-based Infrastructure to Easily Test State-of-the-art DRAM Chips}},
	author       = {Olgun, Ataberk and Hassan, Hasan and Ya{\u{g}}l{\i}k{c}{\i}, A. Giray and Tu{\u{g}}rul, Yahya Can and Orosa, Lois and Luo, Haocong and Patel, Minesh and Ergin O{\u{g}}uz and Mutlu, Onur},
	year         = 2023,
	journal      = {{TCAD}}
}

@misc{AppleRefInc,
	title        = {{About the Security Content of Mac EFI Security Update 2015-001}},
	author       = {{Apple Inc.}},
	year         = 2015,
	howpublished = {\url{https://support.apple.com/en-us/HT204934}}
}

@misc{rh-hp,
	title        = {{HP Moonshot Component Pack Version 2015.05.0}},
	author       = {{Hewlett-Packard Enterprise}},
	year         = 2015,
	howpublished = {\url{http://h17007.www1.hp.com/us/en/enterprise/servers/products/moonshot/component-pack/index.aspx}}
}

@misc{rh-lenovo,
	title        = {{Row Hammer Privilege Escalation}},
	author       = {{Lenovo}},
	year         = 2015,
	howpublished = {\url{https://support.lenovo.com/us/en/product_security/row_hammer}}
}

@misc{greenfield2012throttling,
	title        = {{Throttling Support for Row-Hammer Counters}},
	author       = {Greenfield, Zvika and Levy, Tomer},
	year         = 2016,
	note         = {{U.S.\ Patent 9,251,885}}
}

@article{kim2014architectural,
	title        = {{Architectural Support for Mitigating Row Hammering in DRAM Memories}},
	author       = {Kim, Dae-Hyun and Nair, Prashant J and Qureshi, Moinuddin K},
	year         = 2014,
	journal      = {IEEE CAL}
}

@misc{bains14d,
	title        = {{Distributed Row Hammer Tracking}},
	author       = {Bains, Kuljit S and Halbert, John B},
	year         = 2016,
	howpublished = {{U.S.}\ Patent: 9,299,400}
}

@misc{bains14c,
	title        = {{Method, Apparatus and System for Providing a Memory Refresh}},
	author       = {Bains, K.S. and others},
	year         = 2015,
	howpublished = {{U.S.}\ Patent: 9,030,903}
}

@inproceedings{aweke2016anvil,
	title        = {{ANVIL: Software-Based Protection Against Next-Generation Rowhammer Attacks}},
	author       = {Aweke, Zelalem Birhanu and Yitbarek, Salessawi Ferede and Qiao, Rui and Das, Reetuparna and Hicks, Matthew and Oren, Yossi and Austin, Todd},
	year         = 2016,
	booktitle    = {ASPLOS}
}

@misc{bains-merged,
	title        = {{Row Hammer Refresh Command}},
	author       = {Bains, K. and others},
	year         = 2015,
	howpublished = {{U.S.}\ Patents: 9,117,544 9,236,110 10,210,925}
}

@inproceedings{son2017making,
	title        = {{Making DRAM Stronger Against Row Hammering}},
	author       = {Son, Mungyu and Park, Hyunsun and Ahn, Junwhan and Yoo, Sungjoo},
	year         = 2017,
	booktitle    = {DAC}
}

@inproceedings{seyedzadeh2018cbt,
	title        = {{Mitigating Wordline Crosstalk Using Adaptive Trees of Counters}},
	author       = {S. M. {Seyedzadeh} and A. K. {Jones} and R. {Melhem}},
	year         = 2018,
	booktitle    = {ISCA}
}

@article{irazoqui2016mascat,
	title        = {{MASCAT: Stopping Microarchitectural Attacks Before Execution}},
	author       = {Irazoqui, Gorka and Eisenbarth, Thomas and Sunar, Berk},
	year         = 2016,
	journal      = {IACR Cryptology}
}

@inproceedings{you2019mrloc,
	title        = {{MRLoc: Mitigating Row-Hammering Based on Memory Locality}},
	author       = {You, Jung Min and Yang, Joon-Sung},
	year         = 2019,
	booktitle    = {DAC}
}

@inproceedings{lee2019twice,
	title        = {{TWiCe: Preventing Row-Hammering by Exploiting Time Window Counters}},
	author       = {Lee, Eojin and Kang, Ingab and Lee, Sukhan and Suh, G. Edward and Ahn, Jung Ho},
	year         = 2019,
	booktitle    = {ISCA}
}

@inproceedings{park2020graphene,
	title        = {{Graphene: Strong yet Lightweight Row Hammer Protection}},
	author       = {Park, Yeonhong and Kwon, Woosuk and Lee, Eojin and Ham, Tae Jun and Ahn, Jung Ho and Lee, Jae W},
	year         = 2020,
	booktitle    = {MICRO}
}

@misc{yaglikci2021security,
	title        = {{Security Analysis of the Silver Bullet Technique for RowHammer Prevention}},
	author       = {Ya{\u{g}}l{\i}k{c}{\i}, A. Giray and Kim, Jeremie S. and Devaux, Fabrice and Mutlu, Onur},
	year         = 2021,
	howpublished = {arXiv:2106.07084}
}

@inproceedings{yaglikci2021blockhammer,
	title        = {{BlockHammer: Preventing RowHammer at Low Cost by Blacklisting Rapidly-Accessed DRAM Rows}},
	author       = {Ya{\u{g}}l{\i}k{c}{\i}, A. Giray and Patel, Minesh and Kim, Jeremie S. and Azizibarzoki, Roknoddin and Olgun, Ataberk and Orosa, Lois and Hassan, Hasan and Park, Jisung and Kanellopoullos, Konstantinos and Shahroodi, Taha and Ghose, Saugata and Mutlu, Onur},
	year         = 2021,
	booktitle    = {HPCA}
}

@article{kang2020cattwo,
	title        = {{CAT-TWO: Counter-Based Adaptive Tree, Time Window Optimized for {DRAM} Row-Hammer Prevention}},
	author       = {Ingab Kang and Eojin Lee and Jung Ho Ahn},
	year         = 2020,
	journal      = {{IEEE} Access}
}

@inproceedings{qureshi2022hydra,
	title        = {{Hydra: Enabling Low-Overhead Mitigation of Row-Hammer at Ultra-Low Thresholds via Hybrid Tracking}},
	author       = {Qureshi, Moinuddin K. and Rohan, Aditya and Saileshwar, Gururaj and Nair, Prashant J},
	year         = 2022,
	booktitle    = {ISCA}
}

@inproceedings{saileshwar2022randomized,
	title        = {{Randomized Row-Swap: Mitigating Row Hammer by Breaking Spatial Correlation Between Aggressor and Victim Rows}},
	author       = {Saileshwar, Gururaj and Wang, Bolin and Qureshi, Moinuddin and Nair, Prashant J},
	year         = 2022,
	booktitle    = {ASPLOS}
}

@article{olgun2022pidram,
  author = {Olgun, Ataberk and Luna, Juan Gomez and Kanellopoulos, Konstantinos and Salami, Behzad and Hassan, Hasan and Ergin, Oguz and Mutlu, Onur},
  title = {{PiDRAM: A Holistic End-to-End FPGA-Based Framework for Processing-in-DRAM}},
  year = {2022},
journal={TACO}  
}

@misc{dennard1968dram,
  title={{Field-Effect Transistor Memory}},
  author={Dennard, Robert H.},
  year={1968},
  note={{U.S.}\ Patent 3,387,286}
}

@inproceedings{konoth2018zebram,
	title        = {{ZebRAM: Comprehensive and Compatible Software Protection Against Rowhammer Attacks}},
	author       = {Konoth, Radhesh Krishnan and Oliverio, Marco and Tatar, Andrei and Andriesse, Dennis and Bos, Herbert and Giuffrida, Cristiano and Razavi, Kaveh},
	year         = 2018,
	booktitle    = {OSDI}
}

@inproceedings{vig2018rapid,
	title        = {{Rapid Detection of Rowhammer Attacks Using Dynamic Skewed Hash Tree}},
	author       = {Vig, Saru and Bhattacharya, Sarani and Mukhopadhyay, Debdeep and Lam, Siew-Kei},
	year         = 2018,
	booktitle    = {HASP}
}

@inproceedings{kim2022mithril,
	title        = {{Mithril: Cooperative Row Hammer Protection on Commodity DRAM Leveraging Managed Refresh}},
	author       = {Kim, Michael Jaemin and Park, Jaehyun and Park, Yeonhong and Doh, Wanju and Kim, Namhoon and Ham, Tae Jun and Lee, Jae W and Ahn, Jung Ho},
	year         = 2022,
	booktitle    = {HPCA}
}

@inproceedings{lee2021cryoguard,
	title        = {{CryoGuard: A Near Refresh-Free Robust DRAM Design for Cryogenic Computing}},
	author       = {Lee, Gyu-Hyeon and Na, Seongmin and Byun, Ilkwon and Min, Dongmoon and Kim, Jangwoo},
	year         = 2021,
	booktitle    = {ISCA}
}

@inproceedings{marazzi2022protrr,
	title        = {{REGA: Scalable Rowhammer Mitigation with Refresh-Generating Activations}},
	author       = {Marazzi, Michele and Jattke, Patrick and Solt, Flavien and Razavi, Kaveh},
	year         = 2022,
	booktitle    = {{S\&P}}
}

@inproceedings{zhang2022softtrr,
	title        = {{SoftTRR: Protect Page Tables against Rowhammer Attacks using Software-only Target Row Refresh}},
	author       = {Zhang, Zhi and Cheng, Yueqiang and Wang, Minghua and He, Wei and Wang, Wenhao and Nepal, Surya and Gao, Yansong and Li, Kang and Wang, Zhe and Wu, Chenggang},
	year         = 2022,
	booktitle    = {USENIX ATC}
}

@inproceedings{joardar2022learning,
	title        = {{Learning to Mitigate RowHammer Attacks}},
	author       = {Joardar, Biresh Kumar and Bletsch, Tyler K and Chakrabarty, Krishnendu},
	year         = 2022,
	booktitle    = {DATE}
}

@inproceedings{juffinger2023csi,
	title        = {{CSI: Rowhammer--Cryptographic Security and Integrity against Rowhammer}},
	author       = {Juffinger, Jonas and Lamster, Lukas and Kogler, Andreas and Eichlseder, Maria and Lipp, Moritz and Gruss, Daniel},
	year         = 2023,
	booktitle    = {S\&P}
}

@inproceedings{yuksel2025indram,
    title={{In-DRAM True Random Number Generation Using Simultaneous Multiple-Row Activation: An Experimental Study of Real DRAM Chips}},
    author={Yuksel, Ismail Emir and Olgun, Ataberk and Bostanci, F and Canpolat, Oguzhan and Oliveira, Geraldo F and Sadrosadati, Mohammad and Yaglikci, Abdullah Giray and Mutlu, Onur},
    booktitle={ICCD},
    year={2025}
}

@inproceedings{yuksel2024simultaneous,
    title={{Simultaneous Many-Row Activation in Off-the-Shelf DRAM Chips: Experimental Characterization and Analysis}},
    author={Y{\"u}ksel, Ismail Emir and Tu{\u{g}}rul, Yahya Can and Bostanc{\i}, F Nisa and Oliveira, Geraldo F and Ya{\u{g}}l{\i}k{\c{c}}{\i}, A Giray and Olgun, Ataberk and Soysal, Melina and Luo, Haocong and G{\'o}mez-Luna, Juan and Sadrosadati, Mohammad and others},
    booktitle={DSN},
    year={2024}
}

@inproceedings{yaglikci2022hira,
	title        = {{HiRA: Hidden Row Activation for Reducing Refresh Latency of Off-the-Shelf DRAM Chips}},
	author       = {Ya{\u{g}}l{\i}k{c}{\i}, A Giray and Olgun, Ataberk and Patel, Minesh and Luo, Haocong and Hassan, Hasan and Orosa, Lois and Ergin, O{\u{g}}uz and Mutlu, Onur},
	year         = 2022,
	booktitle    = {MICRO}
}

@inproceedings{yuksel2024functionally,
  title={{Functionally-Complete Boolean Logic in Real DRAM Chips: Experimental Characterization and Analysis}},
  author={Y{\"u}ksel, {\.I}smail Emir and Tu{\u{g}}rul, Yahya Can and Olgun, Ataberk and Bostanc{\i}, F Nisa and Ya{\u{g}}l{\i}k{\c{c}}{\i}, A Giray and Oliveira, Geraldo F and Luo, Haocong and G{\'o}mez-Luna, Juan and Sadrosadati, Mohammad and Mutlu, Onur},
  booktitle={HPCA},
  year={2024}
}

@inproceedings{baek2025marionette,
    title={{Marionette: A RowHammer Attack via Row Coupling}},
    author={Baek, Seungmin and Wi, Minbok and Park, Seonyong and Nam, Hwayong and Kim, Michael Jaemin and Kim, Nam Sung and Ahn, Jung Ho},
    booktitle={ASPLOS},
    year={2025}
}

@article{bepary2022dram,
    title={{DRAM Retention Behavior with Accelerated Aging in Commercial Chips}},
    author={Bepary, Md Kawser and Talukder, Bashir Mohammad Sabquat Bahar and Rahman, Md Tauhidur},
    journal={Applied Sciences},
    year={2022}
}

@inproceedings{talukder2020towards,
    title={{Towards the Avoidance of Counterfeit Memory: Identifying the DRAM Origin}},
    author={Talukder, BMS Bahar and Menon, Vineetha and Ray, Biswajit and Neal, Tempestt and Rahman, Md Tauhidur},
    booktitle={HOST},
    year={2020}
}

@inproceedings{saxena2022aqua,
	title        = {{AQUA: Scalable Rowhammer Mitigation by Quarantining Aggressor Rows at Runtime}},
	author       = {Saxena, Anish and Saileshwar, Gururaj and Nair, Prashant J. and Qureshi, Moinuddin},
	year         = 2022,
	booktitle    = {MICRO}
}

@article{enomoto2022efficient,
	title        = {{Efficient Protection Mechanism for CPU Cache Flush Instruction Based Attacks}},
	author       = {Enomoto, Shuhei and Kuzuno, Hiroki and Yamada, Hiroshi},
	year         = 2022,
	journal      = {IEICE TIS}
}

@inproceedings{manzhosov2022revisiting,
	title        = {{Revisiting Residue Codes for Modern Memories}},
	author       = {Manzhosov, Evgeny and Hastings, Adam and Pancholi, Meghna and Piersma, Ryan and Ziad, Mohamed Tarek Ibn and Sethumadhavan, Simha},
	year         = 2022,
	booktitle    = {MICRO}
}

@inproceedings{orosa2021codic,
  title={{CODIC: A Low-Cost Substrate for Enabling Custom In-DRAM Functionalities and Optimizations}},
  author={Orosa, Lois and Wang, Yaohua and Sadrosadati, Mohammad and Kim, Jeremie S and Patel, Minesh and Puddu, Ivan and Luo, Haocong and Razavi, Kaveh and G{\'o}mez-Luna, Juan and Hassan, Hasan and others},
  booktitle={ISCA},
  year={2021}
}

@inproceedings{ajorpaz2022evax,
	title        = {{EVAX: Towards a Practical, Pro-active \& Adaptive Architecture for High Performance \& Security}},
	author       = {Ajorpaz, Samira Mirbagher and Moghimi, Daniel and Collins, Jeffrey Neal and Pokam, Gilles and Abu-Ghazaleh, Nael and Tullsen, Dean},
	year         = 2022,
	booktitle    = {MICRO}
}

@misc{naseredini2022alarm,
	title        = {{ALARM: Active LeArning of Rowhammer Mitigations}},
	author       = {Naseredini, Amir and Berger, Martin and Sammartino, Matteo and Xiong, Shale},
	year         = 2022,
	howpublished = {\url{https://users.sussex.ac.uk/~mfb21/rh-draft.pdf}}
}

@article{joardar2022machine,
	title        = {{Machine Learning-based Rowhammer Mitigation}},
	author       = {Joardar, Biresh Kumar and Bletsch, Tyler K. and Chakrabarty, Krishnendu},
	year         = 2022,
	journal      = {TCAD}
}

\end{document}